\documentclass{article}
\usepackage{amssymb}
\usepackage{amscd}\usepackage[english]{babel}

\def\T{{\rm T}}
\def\s{{\rm t}}
\def\sc{$\Sigma_c$}

\def\beq{\begin{equation}}
\def\eeq{\end{equation}}
\def\ba{\beq\begin{array}{l}}
\def\ea{\end{array}\eeq}
\def\be{\ba}
\def\ee{\ea}

\def\p{\partial}
\def\l{{\lambda}}
\def\b{{\beta}}
\def\a{{\alpha}}
\def\b{{\beta}}

\def\t{{ \theta}}

\def\cO{\mathcal{O}}

\def\t{\hat t}

\begin{document}

\begin{titlepage}
\centerline{}
\vskip 2in
\begin{center}
\large{\Large{\bf FROM PETROV-EINSTEIN TO NAVIER-STOKES}}
\vskip 0.5in
{Vyacheslav Lysov
and Andrew Strominger}
\vskip 0.3in
{\it Center for the Fundamental Laws of Nature,
Harvard University\\
Cambridge, MA, 02138

}
\end{center}
\vskip 0.5in
\begin{abstract}
We consider a $p$+1-dimensional timelike hypersurface \sc\ embedded with a flat induced metric in a $p$+2-dimensional Einstein geometry.   It is shown that imposing a Petrov type I condition on the hypersurface geometry reduces the degrees of freedom in  the extrinsic curvature of  \sc\ to those of a fluid in \sc. Moreover, expanding around a limit in which the mean curvature of the embedding  diverges, the  leading-order Einstein constraint equations on \sc\ are shown to reduce to the non-linear incompressible Navier-Stokes equation for a fluid moving in \sc.

\end{abstract}

\end{titlepage}

\setcounter{page}{1}
\pagenumbering{arabic}

\tableofcontents

\section{Introduction}
  Recently \cite{bkls,Bredberg:2011jq,Compere:2011dx} the dynamics of horizons  in general relativity was isolated and studied by imposing Dirchlet-like boundary conditions on the induced metric of a surface \sc\ at a small distance $\l$ from the horizon while demanding regularity on the future horizon and no incoming flux across the past horizon. It was shown that in a suitably defined near-horizon $\l\to 0$ limit, the horizon dynamics are governed by the incompressible Navier-Stokes equation.  Put another way, the universal fixed-point behavior of near-horizon scaling in general relativity is the same as that of hydrodynamic scaling in fluid dynamics. An explicit expression for the $\l$ expansion of the Einstein geometry in terms of the Navier-Stokes velocity field $v^i$ and pressure $P$ was constructed. 
  
  As an intriguing aside, it was further noted in \cite{Bredberg:2011jq} that for the four-dimensional case the geometry so constructed is, at least at leading nontrivial order in $\l$, of an algebraically special variety known as restricted Petrov type \cite{petr, exac}.\footnote{  It would be interesting to understand if this algebraic specialty persists to all orders for some choice of higher-order boundary conditions.}  In this paper we turn the logic around and show, in every dimension, that imposing a Petrov type I condition in suitable circumstances reduces the Einstein equation to the Navier-Stokes equation in one lower dimension. Hence regularity on the future horizon and the Petrov type I condition are equivalent (at least) as far as the universal scaling behavior is concerned.\footnote{One way of understanding why there should be such an equivalence  is that for $\l \to 0$ \sc\ approaches the horizon with 
  null normal $\ell$ on the past portion ${\cal H}^-$. Our type I condition is then the vanishing of the Weyl tensor components $\ell^\gamma \ell^\delta C_{\a\gamma  \b \delta}=- \ell^\gamma\nabla_\gamma \sigma _{\a\b} -\theta \sigma_{\a\b}$, where $\sigma_{\a\b}$ is the shear and $\theta$ the expansion of ${\cal H}^-$. This in turn implies the shear in ${\cal H}^-$ vanishes. Equivalently there are no gravity waves passing through ${\cal H}^-$ and  no incoming flux from  the past. We thank Thibault Damour for discussions on this point.}  However, as we shall see, imposing the Petrov condition is mathematically much simpler than imposing regularity.
  
More specifically, we embed an intrinsically flat $p$+1-dimensional timelike hypersurface \sc\ into a $p$+2-dimensional 
solution of Einstein's equation.   We then impose the Petrov type I condition, defined below, with respect to the ingoing and outgoing pair of null vectors whose tangents to \sc\  generate time translations.\footnote{A space is said to be Petrov type I if there is some choice of null vectors with respect to which the Weyl tensor obeys certain identities described below. Due to special features of four dimensions,  every 4D Einstein space is Petrov type I with respect to some null vectors, but not necessarily the ones related to time translations on \sc. Those that are Petrov type I with respect to these null vectors in fact also obey the stronger Petrov type II condition: this is the result quoted in \cite{Bredberg:2011jq}.}   This condition sets to zero a total ${(p+2)(p-1) \over 2}$ components of the Wely tensor.  On \sc\ this  constraint reduces the $(p+1)(p+2)\over 2$ 
components of the extrinsic curvature $K_{ab}$  to $p+2$ unconstrained variables, which may be interpreted as the energy density, velocity field $v^i$ and pressure $P$ of a fluid living on the hypersurface. Simply put, this Petrov condition reduces gravity to a fluid. 
The $p+2$ Einstein constraint equations on \sc\ then become an equation of state and evolution equation for the fluid variables. These highly nonlinear fluid equations are not,  to the best of our knowledge, anything previously encountered in fluid dynamics.  However, we next consider an expansion around a limit where  \sc\  is highly accelerated, i.e. the mean curvature $K$ diverges. At leading order in this expansion, the constraint equations are shown to reduce exactly to the incompressible nonlinear Navier-Stokes equation for $v^i$ and $P$ and the leading-order extrinsic geometry of \sc\ evolves as an incompressible fluid. Hence the Petrov type I condition has the holographic character of relating a theory of gravity in $p+2$ dimensions to a theory without gravity in $p+1$ dimensions. 

In the appendix we describe an an alternate set of boundary conditions on \sc, of possible interest in various contexts discussed therein,  in which the mean curvature $K$ is fixed.  These are shown to differ only at subleading order and also lead to the universal  incompressible Navier-Stokes equation in the near-horizon scaling limit. 

\section{\sc\ hypersurface geometry }
  We wish to consider the ``initial" data on a timelike, $p+1$ dimensional hypersurface in a
  $p+2$-dimensional Einstein space.\footnote{For a nice discussion of the geometrical structures relevant in the current context see
\cite{Gourgoulhon:2005ng, Gourgoulhon:2005ch}.} We take the intrinsic metric to be flat
  \be ds_{p+1}^2=\eta_{ab}dx^adx^b= -(dx^0)^2+\delta_{ij}dx^idx^j
, ~~~~a,b=0,...p,~~~~i,j=1,...p. \ee
 The extrinsic curvature 
$K_{ab}$ is  subject to the $p+1$ ``momentum constraints"
\beq
\p^a (K_{ab}-\eta_{ab}K)=0,\;\;
\eeq  
as well as the ``Hamiltonian constraint'' \beq K_{ab}K^{ab}-K^2=0. \eeq
Satisfying these $p+2$  constraints reduces the $(p+1)(p+2)\over 2$ components of $K_{ab}$  to 
$(p-1)(p+2)\over 2$ locally undetermined  variables. 

Given the bulk Einstein equation
\be G_{\mu\nu}=0,~~~\mu,\nu=0,...p+1,\ee
the Riemann and Weyl tensors are equal and determined on \sc.  
One finds the simple expressions for the projections to \sc
\be \label{swey}
C_{abcd}=K_{ad}K_{bc} -K_{ac}K_{bd}\\
C_{abc(n)} = K_{bc,a}-K_{ac,b}  \\
C_{a(n)b(n)}=K K_{ab}-K_{ac}K^c_b 
\ee 
Here $C_{abc(n)}\equiv C_{abc\mu}n^\mu$ etc. with $n^\mu$ the unit normal to \sc. 

\section{The type I constraint}
In this section we describe the Petrov type I condition in $p+2$ dimensions \cite{Coley:2004jv,coley}.  We first introduce the $p+2$  Newman-Penrose-like vector fields
\beq
\ell^2=k^2=0,\;\; (k,\ell)=1,\;\; (m_i,k)=(m_i,\ell)=0,\;\;  (m_i,m_j)=\delta_{ij}.
\eeq
The spacetime is Petrov type I if for some choice of frame
\be
C_{(\ell)i(\ell)j}\equiv \ell^\mu m_i^\nu \ell^\alpha m_j^\beta C_{\mu\nu\alpha\beta}=0\\
\ee
Now let us choose
\be
m_i=\p_i,\;\; \sqrt{2}\ell = \p_0 -n,\;\; \sqrt{2}k=-\p_0-n
\ee 
where  $n$ is the spacelike unit normal and $\p_i,\p_0$ the tangent vectors to \sc. 
Note that this choice singles out a preferred time coordinate and thus breaks Lorentz invariance of \sc. 
Using (\ref{swey}) the  type I condition for this frame choice is\beq\label{typeI}
2C_{(\ell)i(\ell)j} = (K-K_{00})K_{ij}+2K_{0i}K_{0j} +2K_{ij,0}-K_{ik}K_{~j}^k -K_{0i,j}-K_{0j,i}=0 
\eeq

Since the Weyl tensor is traceless, the type I condition imposes ${p(p+1) \over 2}-1$ 
conditions on the $(p+1)(p+2)\over 2$ components of $K_{ab}$. We may think of it as determining the trace-free part of $K_{ij}$ in terms of $K_{0i},~K_{00}$ and $K$. This leaves 
$p+2$ independent components, which is exactly the number of components of a compressible fluid with a local pressure, energy and momentum density.  The Hamiltonian constraint 
\beq\label{Ham} G_{\mu\nu}n^\mu n^\nu |_{\Sigma_c} =\frac12 (K^2-K_{ab}K^{ab})=0 \eeq
can be viewed as an equation of state relating the pressure and energy density. The $p+1$ momentum constraint equations 
\beq G_{\mu b}n^\mu  |_{\Sigma_c}=\p^aK_{ab}-\p_bK=0, \eeq
where $G_{\mu b}$ denotes the projection of the second index onto \sc, are then the evolution equations for the fluid. 
Hence these $p+2$ constraints eliminate all local freedom on \sc, and reduce it to a boundary value problem on a $p$-dimensional initial spacelike slice of \sc. 

Hence the Petrov condition has a holographic nature: it reduces a theory of gravity to a theory of a fluid without gravity in one less dimension. However, without any further expansion the 
fluid described here has rather exotic dynamical equations. In the next section we will see the dynamics became familiar when expanded around a large mean curvature limit.

\section{  The large mean curvature  expansion}

We now introduce a parameter $\lambda$ into the general fluid solution  and then expand in $\lambda$.  The first step is to define 
\be \label{ppk}
\tau=\lambda x^0
\ee
so that
\be \label{iik}
ds^2  = \eta_{ab}dx^adx^b= -\frac{d\tau^2}{\lambda^2} +dx_i dx^i.\\
\ee
We describe the extrinsic geometry  in terms of the stress tensor $t^a_{~b}$  given in terms of $K^a_{~b}$  by   
\beq\label{dsa}
t^\tau_{~\tau}=K^j_{~j},~~{t}= pK-{p\over2 \lambda},~~\hat t^i_{~j}=-K^i_{~j}-{\rm trace},~~t^\tau_{~i}=-K^\tau_{~i}
\eeq
where by construction $\hat t^i_{~i}=0$. We have separated out, in the definition of $t^a_{~b}$, a constant "pressure" piece which will diverge as $\lambda \to 0$.  
When all other components in (\ref{dsa}) except this diverging  piece vanish, $K^\tau_{~\tau}={1 \over2 \lambda}$ and  \sc\ is then simply the 
hyperbola in the Rindler wedge of Minkowski space 
\beq
ds^2 = -{r}dt^2 +2dt dr+dx_i dx^i\;,
\eeq
located at $r={\lambda}^2$ (note $\tau=\lambda^2 t$). For $\lambda \to 0$ the mean curvature of \sc\ becomes large and it approaches its own future horizon.  Hence the $\lambda\to 0$ limit can be thought of as a kind of near-horizon limit. 

More generally, for finite $\lambda$,  the type I conditions (\ref{typeI}) written in terms of the variables (\ref{dsa}) have the following form 
\beq \label{trless}
\left( t_{~\tau}^{\tau }-\frac2p(t- t_{~\tau}^{\tau })- \frac{1}{\lambda}\right)\t^i_{~j}+\frac{2}{\lambda^2} t^{\tau i} t^{\tau}_ {~j}-\t^i_{~k} \t_{~j}^k  -2\lambda \t^i_{~j,\tau} -\frac{2}{\lambda} \delta^{ki}t^{\tau}_{ ~(k,j)} -\hbox{trace} =0
\eeq
with $i,j$ indexes raised and lowered with $\delta_{ij}$. 
Now we expand in powers of $\lambda$ taking $t^a_{~b} \sim \cO(\lambda^0)  $
or smaller. That is, for the components appearing in (\ref{dsa}) 
\beq
t^a_{~b} = \sum_{k=0}^\infty t^{a(k)}_{~b}\lambda^k.
\eeq
As there is only one term of order ${ 1 \over \lambda^2}$ in equation (\ref{trless}) it immediately implies  that the leading term of ${t^\tau}_j \sim \cO(\lambda)$ and the leading term of $\t^i_{~j}$  is 
\beq
\hat t^{(1)}_{ij} = 2 t^{\tau (1)}_{~i}t^{\tau(1)}_{~j} - 2t^{\tau(1)}_{~(i,j)} -\hbox{trace}.
\eeq
The exact Hamiltonian constraint 
\beq\label{ham}
 (t_{~\tau}^{\tau})^2  -2\frac{ (t^{\tau}_{~i}) ^2}{\lambda^2} + t^i_{~j}t^j_{~i}- \frac{1}{\lambda}t^\tau_{~\tau} - \frac1p t^2 =0\\
\eeq
at leading order fixes $t^\tau_{~\tau}$ as 
\beq
t^{\tau(1)}_{~\tau} =- 2t^{\tau (1)}_{~i}t^{\tau (1)}_{~j} \delta^{ij}.
\eeq
Finally we come to the momentum constraints
\beq \p^at_{ab}=0. \eeq
The time component gives at leading order
\beq \label{ijy}
\p^it^{\tau (1)}_{~i}=0.
\eeq
The space components are at leading order 
\beq\label{iue}
\p_\tau t^{\tau(1)} _{~i}+2t^{\tau(1)} _{~k}\p^k t^{\tau(1)} _{~i} -\p^2t^{\tau(1)} _{~i}+\frac1p\p_it^{(1)}=0.
\eeq
Identifying 
\beq t^{\tau(1)}_{~i}=v_i/2,~~~~t^{(1)}=pP/2,  \eeq
as the velocity and pressure fields, (\ref{ijy}) and (\ref{iue}) become 
\be \p_kv^k=0,\ee
\be \p_\tau v_i+v^k\p_kv_i-\p^2v_i+\p_iP=0 .\ee
This is precisely the incompressible Navier-Stokes system in $p$ space dimensions   \cite{TheoLib:GEN15}.

\section{Acknowledgements}
We are grateful to  L. Andersson, D. Berman, I. Bredberg, D. Christodoulou, G. Compere, T. Damour, G. Gibbons,  J. Hartle,  C. Keeler, 
P. McFadden, G. S. Ng, S. Shenker, K. Skenderis, M. Taylor  and S. T. Yau for illuminating conversations. This work was supported by DOE grant DE-FG0291ER40654 and the Fundamental Laws Initiative at Harvard.

\appendix
\section{Appendix: $K$=constant boundary conditions\\ (with I. Bredberg)}\label{appendix}

In this appendix we consider a modification of the flat``Dirichlet" boundary conditions 
$h_{ab}=\eta_{ab}$ imposed on the hypersurface \sc. In general there is freedom at  higher orders in the choice of boundary conditions: any modification of the metric of order $\lambda$ or smaller will not affect the universal emergence of the incompressible Navier-Stokes equation in the $\lambda \to 0$ scaling limit. The flat boundary conditions have been employed for their simplicity and naturalness.  In this appendix we describe  an
alternate boundary condition for which the metric is only conformally flat and the mean curvature $K$ is fixed to a constant. Roughly speaking this is Neumann rather than 
Dirichlet boundary conditions for the metric conformal factor. 

These constant mean curvature boundary conditions are of interest for several reasons. 
Firstly, constant $K$ hypersurfaces have interesting mathematical properties which have been the subject of much study over the last half century. In the present context they seem particularly appropriate because our expansion parameter is $K^{-1}$. Secondly, in recent generalizations to compact spherical horizons \cite{AndyIrene}, a global obstruction (related to total energy conservation) appears at a subleading order which prevents one from fixing the total area of a spatial cross section of \sc . This obstruction is absent in the constant $K$ formulation here which does allow the area 
to change. 

We take the intrinsic metric of \sc\ to be conformally flat
  \be ds_{p+1}^2=e^{2\rho}\eta_{ab}dx^adx^b=e^{2\rho}( -(dx^0)^2+dx_idx^i), \ee 
  where here and elsewhere $i,j$ indices are raised and lowered with $\delta_{ij}$. Instead of fixing $\rho=0$ as above, we take constant mean curvature 
  \be K=e^{-2\rho}\eta^{ab}K_{ab}={1 \over 2 \lambda} \ee
  It is convenient to describe the remaining components of the extrinsic geometry in terms of the conformally transformed, traceless stress tensor 
\be \T_{ab}=e^{(p-1)\rho} K_{ab}-{e^{(p+1)\rho}\over p+1}\eta_{ab}K, \ee
in terms of which the 
the $p+1$ ``momentum constraints" are
\beq
\p^a \T_{ab} =0,\;\;
\eeq  The conformal factor $\rho$ is then determined from the ``Hamiltonian constraint'' or York equation 
\beq    -2p{\p}_a{\p}^a\rho+p(1-p){\p}_a\rho {\p}^a\rho+e^{-2p\rho}{  \T}_{ab}{  \T}^{ab}-\frac{pe^{2\rho} }{4\lambda^2(p+1)}=0,\eeq
with indices here raised and lowered with $\eta$. The Petrov type I condition for $\sqrt{2}\ell = e^{-\rho}\p_0 -n$ is, instead of (\ref{typeI}) 
\be\label{typeIp}
2e^{2\rho}C_{\ell i\ell j}=
\frac{pe^{-(p-1)\rho}}{2\lambda(p+1)}\T_{ij}+e^{-2p\rho} (\T_{0 i}\T_{0 j}-\T_{00}\T_{ij}-\T_{aj}\T^a_i)\\
- \p_{i} (e^{-p\rho}\T_{ j 0}) - \p_{j} (e^{-p\rho}\T_{ i 0}) +2\p_0 (e^{-p\rho}\T_{ij})  + p\;\p_i\p_j\rho -p \;\p_i\rho \p_j\rho -\hbox{trace}=0
\ee
 To define the the large mean curvature  expansion again take $
\tau=\lambda x^0$ so that
\be \label{iik}
ds^2  =e^{2\rho} \eta_{ab}dx^adx^b=e^{2\rho}\left( -\frac{d\tau^2}{\lambda^2} +dx_i dx^i\right).\\
\ee
and instead of (\ref{dsa})\beq\label{dsap}
\s^\tau_{~\tau}= \T^j_{~j}+\frac{p}{2\lambda(p+1)}e^{(p+1)\rho},~~\hat \s^i_{~j}=-\T^i_{~j}-\hbox{trace} ,~~\s^\tau_{~i}=-\T^\tau_{~i}
\eeq
where by construction $\hat \s^i_{~i}=0$.  For these variables  the type I conditions (\ref{typeIp}) written in terms of the variables (\ref{dsap}) have the following form 
\be \label{trlesst}
-\frac{1}{\lambda}e^{\rho}\hat \s_{ij} +e^{-p\rho} [\frac{2}{\lambda^2}\s_{ ~i}^\tau \s^\tau_{~ j}+\frac{p+2}{p}\s^\tau_{~\tau}\hat \s_{ij}  -\hat \s_{ik} \hat \s^k_j ]-
 \frac{1}{\lambda} [(\p_i - p \p_i\rho )\s_{~ j}^\tau +(\p_j - p\p_j\rho) \s_{~ i}^\tau ]\\  -2\lambda (\p_\tau \hat \s_{ij} -p\p_\tau \rho\; \hat \s_{ij}) 
 + p\;\p_i\p_j\rho -p \;\p_i\rho \p_j\rho-\hbox{trace}=0
\ee
with $i,j$ indexes raised and lowered with $\delta_{ij}$. 
Now we expand in powers of $\lambda$ taking $\s^a_{~b} \sim \cO(\lambda^0) $ We also so take $ \rho \sim \cO(\lambda) $
or smaller so that in the limit we recover a fluid in flat space. That is, for the components appearing in (\ref{dsap})
\beq
\s^a_{~b} = \sum_{k=0}^\infty \s^{a(k)}_{~b}\lambda^k,\;\;\; \rho = \sum_{k=1}^\infty \rho^{(k)} \lambda^k 
\eeq
As there is only one term of order ${ 1 \over \lambda^2}$ in equation (\ref{trlesst}) it immediately implies  that the leading term of ${\s^\tau}_j \sim \cO(\lambda)$ and the leading term of ${\hat \s}^i_{~j}$  is 
\beq
\hat \s^{(1)}_{ij} = 2 \s^{\tau (1)}_{~i}\s^{\tau(1)}_{~j} - 2\s^{\tau(1)}_{~(i,j)} -\hbox{trace}.
\eeq
The exact Hamiltonian constraint 
\beq\label{hamt}
-2p{\p}_a{\p}^a\rho+p(1-p){\p}_a\rho {\p}^a\rho+
e^{-2p\rho}\left[\frac{p+1}{p} (\s^\tau_{~\tau})^2-\frac{1}{\lambda}e^{(p+1)\rho} \s^\tau_{~\tau} -\frac{2}{\lambda^2}\s_{~ i}^\tau \s^{\tau i} +\hat \s_{ij}^2  \right ] =0
\eeq
at leading order fixes $\s^\tau_{~\tau}$ as 
\beq
\s^{\tau(1)}_{~\tau} =- 2\s^{\tau (1)}_{~i}\s^{\tau i(1)} .
\eeq
Finally we come to the momentum constraints
\be \p_a \T^{a}_{~0} =\frac{1}{\lambda} \p^i \s_{~i}^{\tau} +\frac{p}{2}  e^{(p+1)\rho} \p_\tau \rho-\lambda \p_\tau  \s^\tau_{~\tau}=0, \\
\p_a \T^{a }_{~j} =  - \p_\tau \s^\tau_{~j} -\p_i \hat \s^i_{~j} +\frac1p \p_j \s^\tau_{~\tau} -   \frac{1}{2\lambda} e^{(p+1)\rho} \p_j\rho=0. 
\ee
The time component gives at leading order
\beq \label{ijyt}
\p^i\s^{\tau (1)}_{~i}=0.
\eeq
The space components are at leading order 
\beq
\p_j \rho^{(1)}=0
\eeq
and at the next order
\beq\label{iuet}
\p_\tau \s^{\tau(1)} _{~i}+2\s^{\tau(1)} _{~k}\p^k \s^{\tau(1)} _{~i} -\p^2\s^{\tau(1)} _{~i}+\frac{1}{2} \p_j \rho^{(2)}=0.
\eeq
Identifying 
\beq \s^{\tau(1)}_{~i}=v_i/2,~~~~\rho^{(2)}=P,  \eeq
as the velocity and pressure fields, (\ref{ijyt}) and (\ref{iuet})  become 
\be \p_kv^k=0,\ee
\be \p_\tau v_i+v^k\p_kv_i-\p^2v_i+\p_iP=0 .\ee
This again is the incompressible Navier-Stokes system in $p$ space dimensions   \cite{TheoLib:GEN15}.

\bibliography{petrov_ref}{}
\bibliographystyle{utphys}

\end{document}